\documentclass[11pt]{article}
\usepackage{amsmath,amssymb,amsthm,amsxtra,bbm,bbold,bm,epsfig,multirow,longtable}
\usepackage{ulem,subfigure,hyperref,cite}
\usepackage{rotating}

\usepackage[dvipsnames]{xcolor}
\def\thefootnote{\fnsymbol{footnote}} 
\allowdisplaybreaks

\definecolor{darkpink}{RGB}{219, 48, 122}

\newcommand{\Dmq}{\Delta m^2}

\newcommand{\eVq}{\ensuremath{\text{eV}^2}}

\begin{document}

\vspace{0.2cm}

\begin{center} 
{\Large\bf Littlest Modular Seesaw}
\end{center}

\begin{center}
{\bf Ivo de Medeiros Varzielas$^1$}\footnote{Email: \tt ivo.de@udo.edu}
,
{\bf Steve F. King$^2$}\footnote{Email: \tt s.f.king@soton.ac.uk}
and 
{\bf Miguel Levy$^1$}\footnote{Email: \tt miguelplevy@ist.utl.pt}
\\\vspace{5mm}
{$^1$CFTP, Departamento de F\'{\i}sica, Instituto Superior T\'{e}cnico,}\\
Universidade de Lisboa,
Avenida Rovisco Pais 1, 1049 Lisboa, Portugal \\
{$^2$ School of Physics and Astronomy, University of Southampton,\\
Southampton SO17 1BJ, United Kingdom } 
\\
\end{center}

\vspace{1.5cm} 

\begin{abstract} 
We present the first complete model of the Littlest Modular Seesaw, based on two right-handed neutrinos, within the framework of multiple modular symmetries, justifying the use of multiple moduli fields which take their values at 3 specific stabilizers of $\Gamma_4 \simeq S_4$, including a new phenomenological possibility.
Using a semi-analytical approach, we perform a $\chi^2$ analysis of each case and show that good agreement with neutrino oscillation data is obtained, including predictive relations between the leptonic mixing angles and the ratio of light neutrino masses, which non-trivially agree with the experimental values. 
It is noteworthy that in this very predictive setup, the models fit the global fits of the 
experimental data remarkably well, both with and without the
Super-Kamiokande atmospheric data, for both choices of stabilizers. 
By extending the model to include a weighton and the double cover group $\Gamma'_4 \simeq S'_4$,
we are able to also account for the hierarchy of the charged leptons using modular symmetries,
without altering the neutrino predictions.
\end{abstract}

\begin{flushleft}
\hspace{0.8cm} PACS number(s): 14.60.Pq, 11.30.Hv, 12.60.Fr \\
\hspace{0.8cm} Keywords: Lepton flavour mixing, flavour symmetry
\end{flushleft}


\vspace*{\fill}
 \centerline{ \textit{Dedicated to Graham G. Ross}}
\vspace*{\fill}

\def\thefootnote{\arabic{footnote}}
\setcounter{footnote}{0}

\newpage

\section{Introduction}

The mysterious threefold replication of the fermion generations is one important open issue of the Standard Model (SM) at the heart of the flavour problem. The most promising solution are symmetries that relate the generations, known as family symmetries or flavour symmetries. Recent reviews include~\cite{King:2017guk,Xing:2019vks,Feruglio:2019ktm}.

The idea of modular invariance~\cite{Ferrara:1989bc,Ferrara:1989qb} has been suggested as key ingredient in solutions to the flavour problem~\cite{Feruglio:2017spp}. In these promising scenarios, a modular symmetry associated with transformations of a modulus field can lead to very predictive models of flavour. The double covers of the groups have also been used in interesting flavour models~\cite{Novichkov:2020eep}.

Nevertheless, in order to apply the methodology of residual flavour symmetries, it is relevant to consider all their fixed points or stabilizers~\cite{Ding:2019gof,deMedeirosVarzielas:2020kji}: special values for the modulus field where part of the modular transformations are preserved. Furthermore, if multiple residual symmetries are desired, multiple modular symmetries, each with its respective modulus, can be employed - as proposed in~\cite{deMedeirosVarzielas:2019cyj} and expanded upon in~\cite{King:2019vhv, deMedeirosVarzielas:2021pug, deMedeirosVarzielas:2022ihu}.

Modular symmetries can further be exploited to explain the mass hierarchy of the fermions by use of an extra field referred to as a weighton~\cite{King:2020qaj}. While similar to the Froggatt-Nielsen mechanism, the weighton explicitly relies on modular invariance and does not require extra symmetry.

Another mass hierarchy that is puzzling is the lightness of neutrino masses.
Although the type I seesaw mechanism can qualitatively explain the smallness of neutrino masses through the heavy right-handed neutrinos (RHNs), if one doesn't make other assumptions, it contains too many parameters to make any particular predictions for neutrino mass and mixing. The sequential dominance (SD)~\cite{King:1998jw,King:1999cm} of right-handed neutrinos proposes that 
the mass spectrum of heavy Majorana neutrinos is strongly hierarchical, i.e. $M_\text{atm}\ll M_\text{sol}\ll M_\text{dec}$,
where the lightest RHN with mass $M_\text{atm}$ is 
responsible for the atmospheric neutrino mass, that with mass $M_\text{sol}$ gives 
the solar neutrino mass, and a third largely decoupled RHN
gives a suppressed lightest neutrino mass. It leads to an effective two right-handed neutrino (2RHN) model ~\cite{King:1999mb,Frampton:2002qc} with a natural explanation for the physical neutrino mass hierarchy, with normal ordering and the lightest neutrino being approximately massless, $m_1=0$.

A very predictive minimal seesaw model with two right-handed neutrinos and one texture zero is the so-called constrained sequential dominance (CSD) model~\cite{King:2005bj,Antusch:2011ic,King:2013iva,King:2015dvf,King:2016yvg,Ballett:2016yod,King:2018fqh,King:2013xba,King:2013hoa,Bjorkeroth:2014vha}.
The CSD($n$) scheme assumes that the two columns of the Dirac neutrino mass matrix are proportional to $(0,1, -1)$ and $(1, n, 2-n)$ respectively in the RHN diagonal basis, where the parameter $n$ was initially assumed to be a positive integer, but in general may be a real number. For example the CSD($3$) (also called Littlest Seesaw model)~\cite{King:2013iva,King:2015dvf,King:2016yvg,Ballett:2016yod,King:2018fqh}, CSD($4$) models~\cite{King:2013xba,King:2013hoa} and CSD($-1/2$)~\cite{Chen:2019oey} can give rise to phenomenologically viable predictions for lepton mixing parameters and the two neutrino
mass squared differences $\Delta m^2_{21}$ and $\Delta m^2_{31}$,
corresponding to special constrained cases of TM1 lepton mixing. 
As was observed, modular symmetry remarkably suggests CSD($1-\sqrt{6}$) $\approx$ CSD($-1.45$)~\cite{Ding:2019gof,Ding:2021zbg}, although such a model would require multiple moduli and so far there is no complete model of this kind in the literature.

In this paper, we construct the first complete model of the Littlest Modular Seesaw (LMS), 
based on CSD($1-\sqrt{6}$) $\approx$ CSD($-1.45$),
within a consistent framework based on 
multiple modular symmetries. We also propose a new related possibility based on 
CSD($1+\sqrt{6}$) $\approx$ CSD($3.45$), intermediate between CSD($3$) and CSD($4$).
In each case, three $S_4$ modular symmetries are introduced, each with their respective modulus field at a distinct stabilizer,
leading to three separate residual subgroups, thus dispensing with vacuum alignment mechanisms. The result, in the symmetry basis, is a diagonal charged lepton mass matrix and a LMS scenario of a particular kind.
In order to account for the hierarchy of the charged lepton masses, we subsequently introduce a weighton field, where this model is implemented by upgrading the modular symmetries to the respective double covers $S_4'$.
Using a semi-analytical approach, we perform a $\chi^2$ analysis of each case case and show that good agreement with neutrino oscillation data is obtained, for both possible octants of atmospheric angle,
including predictive relations between the leptonic mixing angles and the ratio of light neutrino masses, which non-trivially agree with the experimental values. It is noteworthy that in this very predictive setup, all the models fit the experimental data remarkably well,
depending on the choice of stabilizers and data set, in one case to within approximately $1 \sigma$.

In Section~\ref{sec:symassign} we present the model with the respective fields and their assignments under the multiple modular symmetries. The charged-lepton structure is shown in Section~\ref{sec:charged-leptons}, and the neutrino seesaw matrix is shown in sec~\ref{sec:neutrinos}. Analytical results for the leptonic mixing angles and the neutrino masses are given in Section~\ref{sec:Analytical} and a numerical analysis is done in Section~\ref{sec:Numerical}. We conclude in Section~\ref{sec:conc}. Appendix~\ref{app:weighton} gives two alternative models where the charged-lepton hierarchies are naturally explained by including a weighton.

\section{The Model \label{sec:model}}

\subsection{Symmetries and stabilizers \label{sec:symassign}}

The model we are building features three commuting $S_4$ modular symmetries, which we label as $S_4^A$, $S_4^B$, $S_4^C$. At low energies, due to the VEVs of fields $\Phi_{AC}$ and $\Phi_{BC}$, they are broken down to the diagonal subgroup, as described in~\cite{deMedeirosVarzielas:2019cyj}.  Table~\ref{ta:model} contains the transformation properties (representations and modular weights) under the modular symmetries of the fields and of the relevant modular forms, where we also take usual $SU(2)$ doublets $H_{u,d}$ to transform trivially under all flavour symmetries, and so we omit them from Table~\ref{ta:model}. These assignments are very similar to those used in~\cite{deMedeirosVarzielas:2019cyj}\footnote{We note there is a typo in~\cite{deMedeirosVarzielas:2019cyj}  where RH leptons and the respective modular forms should have primes, as the modular form $Y_\tau(\tau_C)$ (weight 2) only exists as a $3'$.}.

\begin{table}[h]
\centering
\begin{footnotesize}
 \begin{tabular}{| l | c c c c c c|}
\hline \hline
Field & $S_4^A$ & $S_4^B$ & $S_4^C$ & \!$2k_A$\! & \!$2k_B$\! & \!$2k_C$\!\\ 
\hline \hline
$L$ & $\mathbf{1}$ & $\mathbf{1}$ & $\mathbf{3}$ & 0 & 0 & 0\\
$e^c$ & $\mathbf{1}$ & $\mathbf{1}$ & $\mathbf{1}'$ & 0 & 0 & \!$-6$\! \\
$\mu^c$ & $\mathbf{1}$ & $\mathbf{1}$ & $\mathbf{1}'$ & 0 & 0 & \!$-4$\! \\
$\tau^c$ & $\mathbf{1}$ & $\mathbf{1}$ & $\mathbf{1}'$ & 0 & 0 & \!$-2$\! \\
$N_A^c$ & $\mathbf{1'}$ & $\mathbf{1}$ & $\mathbf{1}$ & $-4$ & 0 & 0 \\
$N_B^c$ & $\mathbf{1}$ & $\mathbf{1'}$ & $\mathbf{1}$ & 0 & $-2$ & 0 \\
\hline 
$\Phi_{AC}$ & $\mathbf{3}$ & $\mathbf{1}$ & $\mathbf{3}$ & 0 & 0 & 0 \\
$\Phi_{BC}$ & $\mathbf{1}$ & $\mathbf{3}$ & $\mathbf{3}$ & 0 & 0 & 0 \\
\hline \hline
\end{tabular}
\begin{tabular}{| l | c c c c c c|}
\hline \hline
Yuk/Mass &$S_4^A$ & $S_4^B$ & $S_4^C$ & \!$2k_A$\! & \!$2k_B$\! & \!$2k_C$\!\\
\hline \hline
$Y_e(\tau_C)$ & $\mathbf{1}$ & $\mathbf{1}$ & $\mathbf{3}'$ & 0 & 0 & $6$ \\
$Y_\mu(\tau_C)$ & $\mathbf{1}$ & $\mathbf{1}$ & $\mathbf{3}'$ & 0 & 0 & $4$ \\
$Y_\tau(\tau_C)$ & $\mathbf{1}$ & $\mathbf{1}$ & $\mathbf{3}'$ & 0 & 0 & $2$ \\
$Y_A(\tau_A)$ & $\mathbf{3'}$ & $\mathbf{1}$ & $\mathbf{1}$ & $4$ & 0 & 0 \\
$Y_B(\tau_B)$ & $\mathbf{1}$ & $\mathbf{3'}$ & $\mathbf{1}$ & 0 & $2$ & 0 \\\hline
$M_A(\tau_A)$ & $\mathbf{1}$ & $\mathbf{1}$ & $\mathbf{1}$ & $8$ & 0 & 0 \\
$M_B(\tau_B)$ & $\mathbf{1}$ & $\mathbf{1}$ & $\mathbf{1}$ & 0 & $4$ & 0 
\\
\hline \hline
\end{tabular}
\caption{Transformation properties of fields and modular forms (Yuk/Mass) under the modular symmetries.}
\label{ta:model}
\end{footnotesize}
\end{table}

Our goal is to achieve a CSD(3.45)~\cite{Ding:2019gof} structure from the multiple modular symmetries. To that end, 
the desired directions of the modular forms are obtained for these representations and weights at specific stabilizers~\cite{Ding:2019gof,deMedeirosVarzielas:2020kji, deMedeirosVarzielas:2019cyj}. 
Namely, following the basis of~\cite{deMedeirosVarzielas:2019cyj}, we compute the modular forms\footnote{This choice is not unique, and $\tau_A'=(-3+i)/2$ also gives the same modular form.}:
\begin{equation}
\tau_A = \frac{1}{2}+\frac{i}{2} : \qquad \qquad  Y_{\mathbf{3}'}^{(4)} (\tau_A ) = (0, -1, 1) \, ,
\end{equation}
for one of the Dirac mass matrix columns, and 
\begin{equation}
\label{eq:stab1}
\tau_B = \frac{3}{2}+\frac{i}{2} : \qquad \qquad Y_{ \mathbf{3}'}^{(2)} (\tau_B ) = (1, 1 - \sqrt{6}, 1 + \sqrt{6}) \, ,
\end{equation}
or
\begin{equation}
\label{eq:stab2}
\tau_B = -\frac{1}{2}+\frac{i}{2} : \qquad \qquad Y_{ \mathbf{3}'}^{(2)} (\tau_B ) = (1, 1 + \sqrt{6}, 1 - \sqrt{6}) \, ,
\end{equation}
for the other. These specific modular forms lead to the desired CSD structure. In the same basis, we want to enforce a diagonal structure for the
charged-lepton Yukawa coupling matrices. This can be easily achieved through the weights $2$, $4$, and $6$  modular forms transforming 
as $\mathbf{3'}$, for $\tau_C = \omega$:
\begin{subequations}
\begin{eqnarray}
\label{eq:Ye}
&&Y_{\mathbf{3}'}^{(2)} (\tau_C ) = (0, 1, 0)\\
\tau_C = \omega : \qquad \qquad &&
Y_{ \mathbf{3}'}^{(4)} (\tau_C ) = (0, 0, 1)\\
&&Y_{ \mathbf{3}'}^{(6)} (\tau_C ) = (1, 0, 0)
\label{eq:Ytau}
\end{eqnarray}
\end{subequations}
A subtlety should be noted here. Indeed, for weight 6, there are two independent $\mathbf{3'}$ modular forms, which could spoil the 
diagonal arrangement of the charged-leptons. Nevertheless, for $\tau = \omega$, one of them vanishes, introducing no further parameters.

In Appendix~\ref{app:stabs} it is shown that $\tau_A$ and $\tau'_A$ are stabilisers of $U$, and that $\tau_B$ (either version) is a stabiliser of $SU$ in our chosen basis. It is also shown that the respective modular forms we are using are eigenvectors of the $\mathbf{3'}$ representation matrices.

For clarity, we note that the basis in which the modular forms are computed in the present work follows reference~\cite{deMedeirosVarzielas:2019cyj}, which is different from~\cite{Ding:2019gof}. To be precise, although the $S_4$ basis used here and~\cite{deMedeirosVarzielas:2019cyj} is the same as that in~\cite{Ding:2019gof},  the basis of modular generators
is different, and hence the modular forms differ also. However the physics should be and is basis independent, and indeed the
Yukawa alignments shown above can be achieved for different values of the modulus field in the two different bases.
It useful to present the different stabilisers for both cases which lead to the desired modular forms, which is shown in Table~\ref{tab:Dictionary}\footnote{Note that with multiple moduli, transforming under a diagonal $S_4$ subgroup,
it is meaningful to have fixed points outside the fundamental domain.}.
\begin{table}[h!]
\begin{footnotesize}
\begin{center}
\begin{tabular}{|c|ccc|}
\hline\hline
\rule{0pt}{6mm}\ignorespaces
      & $Y^{(4)}_\mathbf{3'}\left(\tau\right) = \begin{pmatrix} 0&  -1& 1 \end{pmatrix}$ & $Y^{(2)}_\mathbf{3'}\left( \tau \right) = \begin{pmatrix} 1 & 1 - \sqrt{6} & 1 + \sqrt{6} \end{pmatrix}$ & $Y^{(2)}_\mathbf{3'}\left( \tau \right) = \begin{pmatrix} 1 & 1 + \sqrt{6} & 1 - \sqrt{6} \end{pmatrix}$ \\[2mm]
\hline
\rule{0pt}{8mm}\ignorespaces
Basis 1 & $\tau = \dfrac{1+i}{2}$, $\tau =  \dfrac{-3+i}{2}$ & $\tau = \dfrac{3+i}{2}$ & $\tau = \dfrac{-1+i}{2}$\\[3mm]
Basis 2 & $\tau = 2+i$, $\tau = \dfrac{-2+i}{5}$ & $\tau = \dfrac{-8+i}{13}$ & $\tau = i$\\[3mm]
\hline\hline
\end{tabular}
\caption{\label{tab:Dictionary} Relevant stabilisers to obtain the desired modular forms to achieve either a CSD(3.45) or a CSD(-1.45) model, both for basis 1 (used throughout this paper), and basis 2 used in \cite{Ding:2019gof}. For both cases, the charged leptons are at the left cusp: $\tau_C = \omega$. Note that the convention of $\mathbf{3}$ and $\mathbf{3'}$ is exchanged.}
\end{center}
\end{footnotesize}
\end{table}

\subsection{Charged leptons \label{sec:charged-leptons}}

With the fields and assignments of the previous subsection, we write the respective lepton sector superpotential as
\begin{eqnarray}\label{eq:superpot}
w_\ell &=&
\frac{1}{\Lambda}\left[L \Phi_{AC} Y_A(\tau_A) N_A^c + L \Phi_{BC} Y_B(\tau_B) N_B^c \right] H_u \nonumber \\
&&+ \left[ L Y_e(\tau_C) e^c + L Y_\mu(\tau_C) \mu^c + L Y_\tau(\tau_C) \tau^c \right] H_d \\
&&+ \frac{1}{2} M_A(\tau_A) N_A^c N_A^c + \frac{1}{2} M_B(\tau_B) N_B^c N_B^c + M_{AB}(\tau_A,\tau_B) N_A^c N_B^c  \nonumber \, .
\end{eqnarray}

Expanding the superpotential of Eq.~\eqref{eq:superpot}, we can find the mass matrices for the fields after the EWSB.
Due to the nature of the $S_4$ tensor products in our chosen basis, and the particular structure chosen for the bi-triplets VEVs, 
the $\mathbf{3}\otimes\mathbf{3}$ tensor products are non-diagonal:
\begin{eqnarray}
\left( \mathbf{a} \otimes \mathbf{b} \right)_\mathbf{1} &=& a_1 b_1 + a_2 b_3 + a_3 b_2, \\
\left( \mathbf{a} \otimes \left<\Phi\right> \otimes \mathbf{b} \right)_\mathbf{1} &\propto&  a_1 b_1 + a_2 b_3 + a_3 b_2. 
\end{eqnarray}

Hence, the charged-lepton mass matrix is simply given by
\begin{equation}
M_l = v_d 
\begin{pmatrix}
 \left(Y_e\right)_{1} & \left(Y_\mu\right)_{1} & \left(Y_\tau\right)_{1} \\
 \left(Y_e\right)_{3} & \left(Y_\mu\right)_{3} & \left(Y_\tau\right)_{3} \\
 \left(Y_e\right)_{2} & \left(Y_\mu\right)_{2} & \left(Y_\tau\right)_{2} 
\end{pmatrix},
\end{equation}
where we omit the $\tau_c$ dependency for clarity, and $v_d$ stands for $\left<H_d\right>$. 
Plugging in the specific shapes of the modular forms given in Eqs.~\eqref{eq:Ye}-\eqref{eq:Ytau} 
we arrive at a diagonal charged-lepton mass matrix when $\tau_C=\omega$:
\begin{equation}
M_l = v_d 
\begin{pmatrix}
 y_e & 0 & 0 \\
 0 & y_\mu & 0 \\
 0 & 0 & y_\tau 
\end{pmatrix}.
\end{equation}

In this model, the hierarchical masses of the charged-leptons are not addressed. In order to naturally deal with this issue, we present two modifications of this model in the Appendix~\ref{app:weighton}, where a weighton is responsible for the hierarchy of the masses, without affecting the remaining predictions of the model.

\subsection{Neutrinos \label{sec:neutrinos}}

We now turn to the Majorana mass terms for the neutrinos, $N_A^C$ and $N_B^C$.
From Table~\ref{ta:model}, we see that $N_A^C N_{A}^C$ as well as $N_B^C N_{B}^C$ are $S_4^A \times S_4^B \times S_4^C$ singlets.
As such, we just need to cancel out the weight with a singlet Yukawa form. From~\cite{Novichkov:2020eep, Novichkov:2018ovf} \footnote{Although we use a different basis, the assignments of the representations are identical, as can be seen by the weight 2 modular forms. Furthermore, we have explicitly checked that the tensor product of $\left(Y_\mathbf{3'}^{(2)} \otimes Y_\mathbf{3'}^{(2)}\right)_\mathbf{1}$ does not vanish for the relevant $\tau_A$ nor any of $\tau_B$. This ensures a non-zero $M_A$ and $M_B$.} we see that the Yukawa modular forms of 
weight 4 do have a singlet representation, needed for the $M_A(\tau_A)$ term. Due to the properties of the modular terms, this implies that there is also 
a singlet modular form of weight 8, required for $M_B(\tau_B)$. Conversely, as $N_A^C N_B^C$ transforms non-trivially under both $S_4^{A}$ and
 under $S_4^B$, there are no one-dimensional modular forms of weight 2 and the respective term is forbidden by the symmetries, and the RH neutrino mass matrix is diagonal:
\begin{equation}
M_R = \begin{pmatrix} M_A(\tau_A) & 0 \\ 0 & M_B(\tau_B) \end{pmatrix}.
\end{equation}

Finally, we need to check the shape of the Dirac mass matrices. Given the VEVs for the bi-triplets $\Phi_{AC}, \Phi_{BC}$, 
the tensor products after SSB will mimic those of the usual $S_4$ (the diagonal $S_4$ preserved by the bi-triplets symmetry breaking), as explained in~\cite{deMedeirosVarzielas:2019cyj, King:2019vhv, deMedeirosVarzielas:2021pug, deMedeirosVarzielas:2022ihu}. This feature is preserved also in the weighton versions of the model, that are using $S'_4$.
The Dirac mass matrix is then given by:
\begin{equation}
M_D = v_u 
\begin{pmatrix}
\left(Y_A\right)_1 & \left(Y_B\right)_1 \\
\left(Y_A\right)_3 & \left(Y_B\right)_3 \\
\left(Y_A\right)_2& \left(Y_B\right)_2  
\end{pmatrix},
\end{equation}
where, as usual, $v_u$ denotes the $H_u$ VEV, and the $2\times3$ structure comes from the CSD with just two RH neutrinos.
Choosing specific stabilisers for the two remaining moduli fields, we can achieve a new CSD(3.45) structure with 
$n=1+\sqrt6$:
\begin{equation}
\label{eq:DiracMat}
M_D = v_u 
\begin{pmatrix}
0&   b\\
a&  b\left(1+ \sqrt{6}\right) \\
-a & b\left(1-\sqrt{6}\right) 
\end{pmatrix}, \qquad \tau_A = -\frac{3}{2}+\frac{i}{2}, \quad \tau_B = \frac{3}{2}+\frac{i}{2}.
\end{equation}
We can similarly achieve the case CSD($-1.45)$ with $n=1-\sqrt6$ already discussed in~\cite{Ding:2019gof}:
\begin{equation}
\label{eq:DiracMat}
M_D = v_u 
\begin{pmatrix}
0&   b\\
a&  b\left(1- \sqrt{6}\right) \\
-a & b\left(1+\sqrt{6}\right) 
\end{pmatrix}, \qquad \tau_A = -\frac{3}{2}+\frac{i}{2}, \quad \tau_B = -\frac{1}{2}+\frac{i}{2}.
\end{equation}

The type-I seesaw mechanism will lead to an effective mass matrix for the light neutrinos:
\begin{equation}
m_\nu = M_D \cdot M_R^{-1} \cdot M_D^T = v_u^2 
\begingroup
\setlength\arraycolsep{15pt}
\begin{pmatrix}  
\dfrac{b^2}{M_B} & \dfrac{b^2 n}{M_B} &  \dfrac{b^2(2-n)}{M_B} \\[12pt]
. & \dfrac{a^2}{M_A} + \dfrac{b^2 n^2}{M_B} & -\dfrac{a^2}{M_A} + \dfrac{b^2n(2-n)}{M_B} \\[12pt] 
. & . & \dfrac{a^2}{M_A} + \dfrac{b^2(2-n)^2}{M_B}
\end{pmatrix},
\endgroup
\label{eq:mnu_mee}
\end{equation}
where $n=1+\sqrt{6} \approx 3.45$ or $n=1-\sqrt{6} \approx -1.45$.

\subsection{Analytic results \label{sec:Analytical}}

The effective mass matrix for the light neutrinos can be split into two contributions, 
\begin{equation}
m_\nu = \frac{v_u^2}{M_A} |a|^2
\begin{pmatrix} 0 & 0 & 0 \\ 0 & 1 & -1 \\ 0 & -1 & 1 \end{pmatrix} 
+ \frac{v_u^2}{M_B} |b|^2 e^{i \beta} 
\begin{pmatrix} 1 & n & 2-n \\ n & n^2 &  n (2-n) \\ 2-n & n (2-n) & (2-n)^2 \end{pmatrix}.
\end{equation}
It is worth noting that the above neutrino mass matrix in the diagonal charged lepton mass basis
is determined effectively by two real parameters, 
$m_a = v_u^2 \frac{\lvert a\rvert ^2}{M_A}\,, \quad m_b = v_u^2 \frac{\lvert b\rvert ^2}{M_B}$, one phase 
$\beta$ and a discrete choice of $n=1\pm \sqrt{6}$.
For a given choice of $n$, the remaining three real parameters determine all the parameters in the neutrino sector,
namely all the neutrino masses and the entire PMNS matrix.

These two terms above can be simultaneously block-diagonalized by the following Tri-bimaximal mixing matrix, 
\begin{equation}
\mathcal{U}_{\rm TBM} = \begin{pmatrix} - \sqrt{\frac{2}{3}} & \sqrt{\frac{1}{3}} & 0 \\ 
\sqrt{\frac{1}{6}} & \sqrt{\frac{1}{3}} & \sqrt{\frac{1}{2}} \\
\sqrt{\frac{1}{6}} & \sqrt{\frac{1}{3}} & -\sqrt{\frac{1}{2}}
\end{pmatrix},
\end{equation}
leading to 
\begin{equation}
m_\nu' = \mathcal{U}^T_{\rm TBM} \cdot m_\nu \cdot \mathcal{U}_{\rm TBM} = \frac{v_u^2}{M_A} |a|^2
 \begin{pmatrix} 0 &0 & 0 \\ 0 & 0 & 0 \\ 0 & 0 & 2 \end{pmatrix} + 
 \frac{v_u^2}{M_B} |b|^2 e^{i \beta}
 \begin{pmatrix} 0 & 0 & 0 \\ 0 & 3 & \sqrt{6} (n-1) \\ 0 & \sqrt{6}(n-1) & 2 (n-1)^2 \end{pmatrix}.
 \end{equation}
We diagonalize the remaining $(2,2)$ block through the matrix
\begin{equation}
\mathcal{U}_\alpha = \begin{pmatrix} 1 & 0 & 0 \\ 0 & c_\alpha &  e^{i \gamma} s_\alpha \\ 0 & - e^{-i \gamma}s_\alpha & c_\alpha \end{pmatrix}, 
\end{equation}
such that 
\begin{equation}
\mathcal{U}_\alpha^T \cdot m_\nu' \cdot \mathcal{U}_\alpha = \text{diag}(0, m_1, m_2).
\end{equation}
To ensure that $m_1, m_2$ are real and positive, we use the phase matrix, $P_\nu$:
\begin{equation}
( U_{\rm TBM} \,  U_\alpha \, P_\nu )^T \, m_\nu \, ( U_{\rm TBM} \,  U_\alpha \, P_\nu ) = \text{diag}(0, \lvert m_1 \rvert, \lvert m_2 \rvert ), 
\end{equation}
with 
\begin{equation}
U_\nu \equiv ( U_{\rm TBM} \,  U_\alpha \, P_\nu ) =  
 \begin{pmatrix}
-\sqrt{\dfrac{2}{3}} &  \dfrac{c_\alpha}{\sqrt{3}}  & e^{i \gamma} \dfrac{s_\alpha}{\sqrt{3}} \\
\sqrt{\dfrac{1}{6}} &    \dfrac{c_\alpha}{\sqrt{3}} - e^{-i\gamma}  \dfrac{s_\alpha}{\sqrt{2}} &  \dfrac{c_\alpha}{\sqrt{2}} + e^{i\gamma} \dfrac{s_\alpha}{\sqrt{3}} \\
\sqrt{\dfrac{1}{6}} &    \dfrac{c_\alpha}{\sqrt{3}}+e^{-i\gamma}   \dfrac{s_\alpha}{\sqrt{2}} & -\dfrac{c_\alpha}{\sqrt{2}} + e^{i\gamma} \dfrac{s_\alpha}{\sqrt{3}}
\end{pmatrix} \cdot
\begin{pmatrix} 1 & 0 & 0 \\ 0 & e^{i \phi_2} & 0 \\ 0 & 0 & e^{i\phi_3} \end{pmatrix}.
\end{equation}

As this is effectively a $2\times2$ diagonalization, it is possible to find analytical relations for $\alpha$. 
Namely, by requiring a vanishing $\left(U_\alpha^T m_\nu'U_\alpha\right)_{23}$ element we find~\cite{King:2015dvf}:
\begin{eqnarray}
t\equiv \tan2\alpha &=& \dfrac{2 y }{z \cos\left(\varphi-\gamma \right) -x \cos\gamma}, \\
\tan\gamma &=& \dfrac{ z \sin\varphi}{x+z\cos\varphi}, \qquad \qquad {\rm with  } \quad \varphi = \phi_z-\beta , \label{eq:varphi}
\end{eqnarray}
where we defined
\begin{equation}
\label{mnublock}
m_\nu' = \begin{pmatrix}
0 & 0 &0 \\
0 & x e^{i \beta} & y e^{i\beta} \\
0 & y e^{i\beta} & z e^{i \phi_z}
\end{pmatrix},
\end{equation}
with 
\begin{equation}
 x = 3 m_b\,, \quad  y = \sqrt{6}(n-1)m_b\,, \quad z = \lvert 2(m_a + e^{i\beta} (n-1)^2 m_b)\rvert\,, \quad m_a = v_u^2 \frac{\lvert a\rvert ^2}{M_A}\,, \quad m_b = v_u^2 \frac{\lvert b\rvert ^2}{M_B}.
\label{eq:mb_def} 
\end{equation}

To relate this to the PMNS matrix in its standard parametrization, we must also take into account the charged-lepton rotation. 
In our specific realisation, the modular representations of the charged-leptons were chosen in such a way that its mass matrix is already diagonal.
As such, the LH rotation is, in general, a diagonal phase matrix
\begin{equation}
U_\ell = \begin{pmatrix} e^{i\delta_e} & 0 & 0 \\ 0 & e^{i \delta_\mu} & 0 \\ 0 & 0 & e^{i\delta_\tau} \end{pmatrix},
\end{equation} 
which can be used to match the standard parametrization~\cite{Workman:2022ynf}\footnote{Indeed, the RH fields rotate away the possible phases of $M_l$ and, as such, when we write down $m_\nu$ we are already in a basis where $M_l$ is diagonal and positive.
The LH rotation was used to enforce the reality of $a$. In general, this won't be the basis where the light neutrino masses are real. $U_l$ is then required to rotate into the standard parametrization basis.}:
\begin{equation}
\label{eq:STDP}
U_{\rm PMNS} =  \begin{pmatrix}
c_{12} c_{13} & s_{12} c_{13} & s_{13}e^{-i\delta} \\
-s_{12} c_{23} -c_{12}s_{13}s_{23}e^{i\delta} & c_{12}c_{23} - s_{12} s_{13} s_{23} e^{i\delta} & c_{13} s_{23} \\
s_{12} s_{23} -c_{12} s_{13} c_{23} e^{i\delta} & -c_{12} s_{23} - s_{12} s_{13} c_{23} e^{i\delta} & c_{13} c_{23}
\end{pmatrix} \cdot \begin{pmatrix} e^{i\eta_1} & 0 & 0 \\ 0 & e^{i \eta_2} & 0 \\ 0 & 0 & 1 \end{pmatrix}, 
\end{equation}
which has the measured mixing angles and CP-violating phase, and $s_{ij}$ ($c_{ij}$) denotes $\sin\theta_{ij}$ ($\cos\theta_{ij}$).

Now, we can relate our Unitary matrix $U_\nu$ to $U_{\rm PMNS}$ and find out the relations between the measured neutrino data, and our model's parameters.
The resulting relations are 
\begin{eqnarray}
\sin\theta_{13} &=& \frac{\sin\alpha}{\sqrt{3}} = \frac{1}{\sqrt{6}} \sqrt{ 1-\sqrt{\dfrac{1}{1+t^2}}}, \\
\tan\theta_{12} &=& \frac{ \cos\alpha}{\sqrt{2}} = \frac{1}{\sqrt{2}} \sqrt{ 1 - 3\sin^2\theta_{13}}, \\
\tan\theta_{23} &=& \dfrac{\lvert 1+\epsilon_{\alpha}\rvert}{\lvert 1-\epsilon_\alpha\rvert} \label{eq:theta23},
\end{eqnarray}
where
\begin{equation}
\epsilon_\alpha = \sqrt{\dfrac{2}{3}} e^{i\gamma} \tan\alpha = \sqrt{\dfrac{2}{3}} e^{i \gamma} \frac{\sqrt{1+t^2}-1}{t}.
\end{equation}

Note that the mixing angles depend only on two parameters, with $\theta_{13}$ and $\theta_{12}$ depending only on $t$. Since the mixing is unaffected by an overall factor, we can factorise $m_b$ in Eq.~\eqref{mnublock}, 
leading to
\begin{equation}
m_\nu' = m_b \begin{pmatrix} 0 & 0 & 0 \\ 0 & x' e^{i \beta} & y' e^{i \beta} \\ 0 & y' e^{i\beta} & z' e^{i\phi_z} \end{pmatrix},
\end{equation}
where 
\begin{eqnarray}
\label{eq:params1}
x' = 3, \qquad y' = \sqrt{6}(n-1), \qquad && z' = \left\lvert 2 \left(\frac{1}{r} + e^{i\beta}(n-1)^2\right)\right\rvert , \\
\phi_z = \text{arg}\left(\frac{1}{r} + e^{i\beta}(n-1)^2\right) , &&\qquad r = \frac{m_b}{m_a} ,\label{eq:paramslast}
\end{eqnarray}
where we note how $\phi_z$ and $z'$ depend on $r$ and $\beta$. For fixed $n$, the mixing angles themselves will depend solely on $r$ and $\beta$.

To obtain the neutrinos masses, we proceed as in~\cite{King:2015dvf} by taking the trace and determinant of the hermitian combination $H'_\nu = {m'_\nu}^\dagger m'_\nu$, and equating it to the sum and product of the squared masses, respectively.
Given that the LS paradigm forcibly leads to a massless light neutrino and thus, to Normal Ordering, the obtained masses can be readily equated to the $\Delta m_{21}^2$ and $\Delta m^2_{31}$ observables. Defining the combinations of parameters, that depend on those of Eqs.~\eqref{eq:varphi} and \eqref{eq:params1}-\eqref{eq:paramslast}, 
\begin{eqnarray}
\Sigma &\equiv& \frac{m^2_b}{2} \left( {x'}^2 + 2 {y'}^2 + {z'}^2 \right),\\
\delta M &\equiv& \frac{m^2_b}{2} \sqrt{ {x'}^2( 4 {y'}^2-2{z'}^2)+{x'}^4 + 8 {x'}{y'}^2{z'}\cos\varphi + 4 {y'}^2 {z'}^2 +{z'}^4} ,
\end{eqnarray}
then
\begin{eqnarray}
\Delta m_{21}^2 = m_2^2 &=& \Sigma - \delta M, \\
\Delta m_{31}^2 = m_3^2 &=& \Sigma + \delta M ,
\end{eqnarray}
which are functions of $r$ and $\beta$, and with the overall factor given by $m_b$, which cancels out in the ratio.
As such, $\Delta m^2_{21}/\Delta m^2_{31}$, the 3 mixing angles, and the CP-phase are all functions of just two effective parameters.

The CP-phase of the PMNS matrix, as well as the physical Majorana phase (since there is one massless neutrino, only $\eta_2$ of Eq.~\eqref{eq:STDP} is physical\footnote{This is made clear when computing $m_{ee}$. Alternatively, we can always rotate $\nu_1$ to absorb $\eta_1$, but this will not influence the second and third columns.}) can be extracted through careful combinations of the elements~\cite{Antusch:2003kp}, and lead to
\begin{eqnarray}
\label{eq:delta}
\delta &=& - \text{arg}\left(\text{sign}(t) e^{i\beta} \left( 4\left( \sqrt{t^2+1} -1 \right) + \left(-2+3e^{2i\gamma}\right)t^2\right)\right) , \\
\eta_2 &=&  \left( -\gamma - \delta - \left(\phi_3 -\phi_2\right) \right).
\end{eqnarray}

\subsection{Numerical analysis \label{sec:Numerical}}
\begin{figure}[h]
\centering
\includegraphics[width=0.8\linewidth]{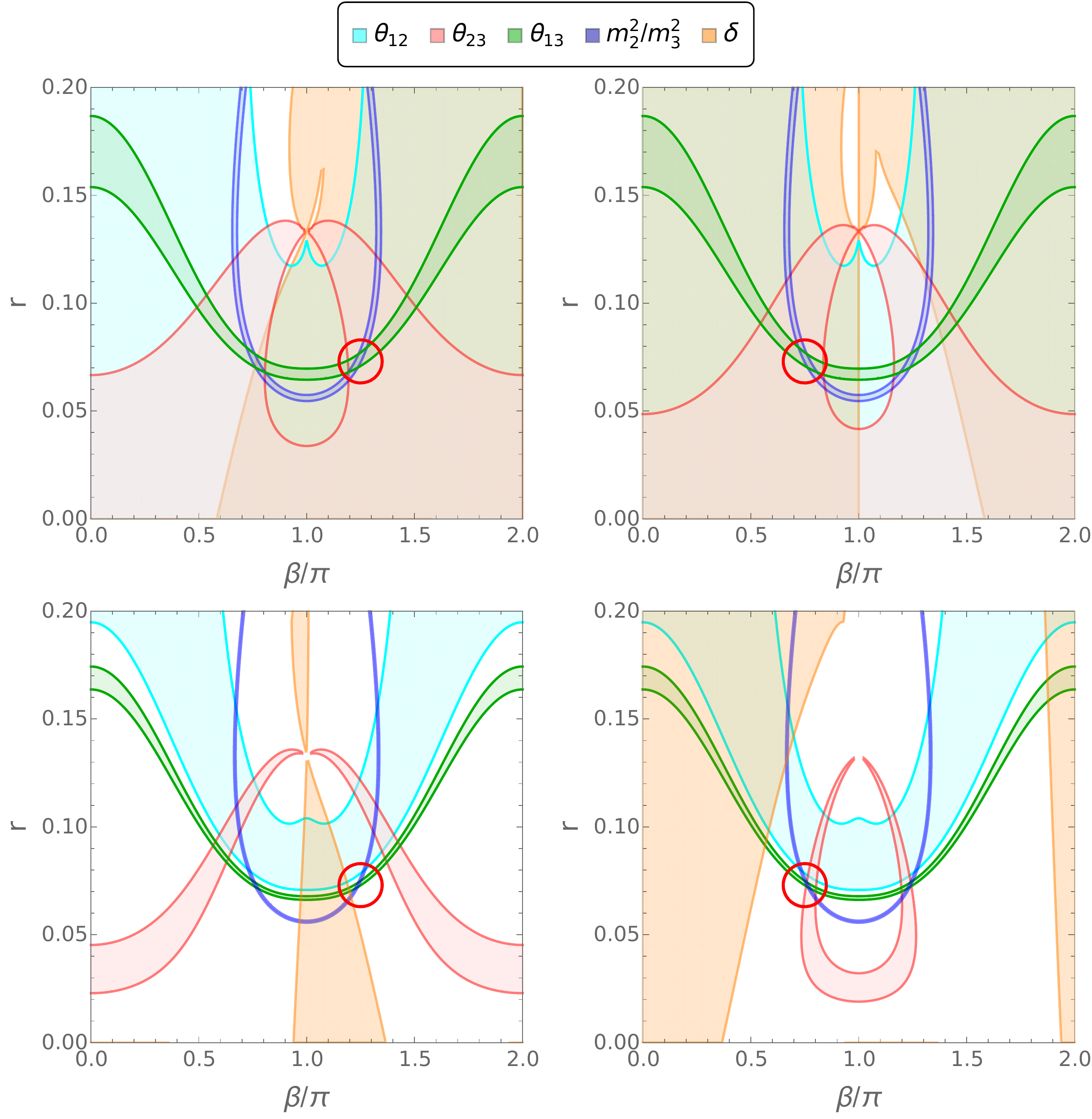}
\caption{\label{fig:WithOut} Allowed $3\sigma$ (\textbf{top}) and $1\sigma$ (\textbf{bottom}) experimental ranges in the $(r, \beta)$ plane using NuFit~5.1 values without SK atmospheric data for the $n=1+\sqrt{6}$ case (\textbf{left}) and for the $n=1-\sqrt{6}$ case (\textbf{right}). The red circle indicates the best fit region.}
\end{figure}

\begin{figure}[h]
\centering
\includegraphics[width=0.8\linewidth]{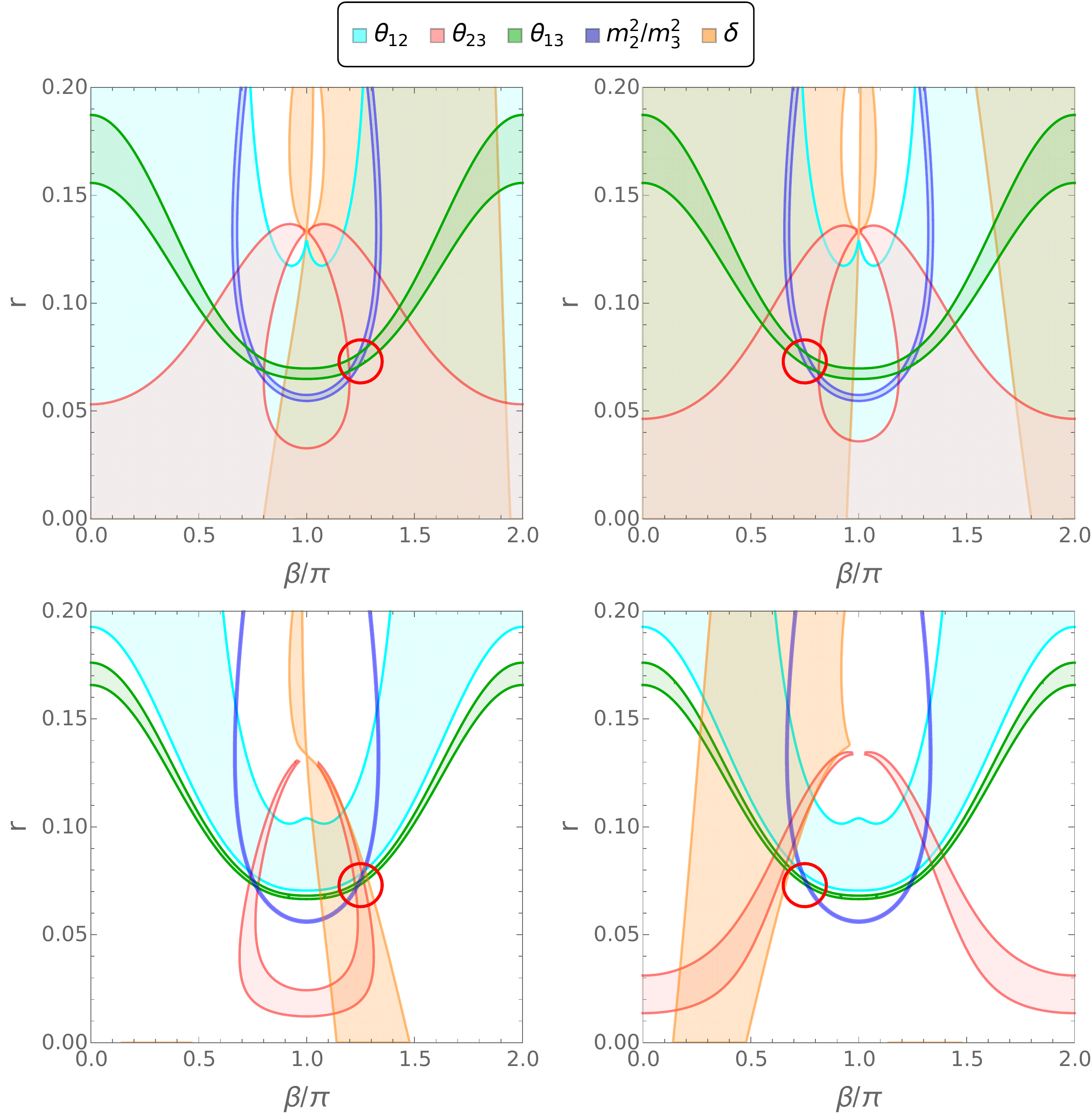}
\caption{\label{fig:With} As in Figure~\ref{fig:WithOut} but using the NuFit~5.1 values with SK atmospheric data. \textbf{Left:} $n=1+\sqrt{6}$, \textbf{right:} $n=1-\sqrt{6}$, \textbf{top:} $3\sigma$, \textbf{bottom:} $1\sigma$.}
\end{figure}

Using the analytical expressions, we plot the allowed experimental ranges for the lepton mixing parameters in the $(r, \beta)$ plane. We present both the case where $\tau_B=(3+i)/2$ and $\tau_B=(-1+i)/2$, corresponding to the modular forms of Eqs~\eqref{eq:stab1} and \eqref{eq:stab2}. The results shown correspond to the NuFit~5.1 values~\cite{Esteban:2020cvm, nufit} without SK atmospheric data in Figure~\ref{fig:WithOut} and with SK atmospheric data in Figure~\ref{fig:With}.
We reproduce the ranges used in Table~\ref{ta:NuFit51}.
In both Figures, the top row displays the $3 \sigma$ ranges, the bottom row the $1 \sigma$ ranges, the left column the $n=1+\sqrt6$ case and the right column the $n=1-\sqrt6$ case \footnote{The results for $n=1-\sqrt6$ match the results of~\cite{Ding:2019gof}, as expected.}.

\begin{table}[h]
\centering
  \begin{footnotesize}
    \begin{tabular}{c|l|cc|cc}
      \hline\hline
      \multirow{11}{*}{\begin{sideways}\hspace*{-7em}Normal Ordering\end{sideways}} &
      & \multicolumn{2}{c|}{without SK atmospheric data}
      & \multicolumn{2}{c}{with SK atmospheric data}
      \\
      \cline{3-6}
      && bfp $\pm 1\sigma$ & $3\sigma$ range
      & bfp $\pm 1\sigma$ & $3\sigma$ range
      \\
      \cline{2-6}
      \rule{0pt}{4mm}\ignorespaces
      & $\sin^2\theta_{12}$
      & $0.304_{-0.012}^{+0.013}$ & $0.269 \to 0.343$
      & $0.304_{-0.012}^{+0.012}$ & $0.269 \to 0.343$
      \\[1mm]
      & $\theta_{12}/^\circ$
      & $33.44_{-0.74}^{+0.77}$ & $31.27 \to 35.86$
      & $33.45_{-0.75}^{+0.77}$ & $31.27 \to 35.87$
      \\[3mm]
      & $\sin^2\theta_{23}$
      & $0.573_{-0.023}^{+0.018}$ & $0.405 \to 0.620$
      & $0.450_{-0.016}^{+0.019}$ & $0.408 \to 0.603$
      \\[1mm]
      & $\theta_{23}/^\circ$
      & $49.2_{-1.3}^{+1.0}$ & $39.5 \to 52.0$
      & $42.1_{-0.9}^{+1.1}$ & $39.7 \to 50.9$
      \\[3mm]
      & $\sin^2\theta_{13}$
      & $0.02220_{-0.00062}^{+0.00068}$ & $0.02034 \to 0.02430$
      & $0.02240_{-0.00062}^{+0.00062}$ & $0.02060 \to 0.02435$
      \\[1mm]
      & $\theta_{13}/^\circ$
      & $8.57_{-0.12}^{+0.13}$ & $8.20 \to 8.97$
      & $8.62_{-0.12}^{+0.12}$ & $8.25 \to 8.98$
      \\[3mm]
      & $\delta/^\circ$
      & $194_{-25}^{+52}$ & $105 \to 405$
      & $230_{-25}^{+36}$ & $144 \to 350$
      \\[3mm]
      & $\dfrac{\Dmq_{21}}{10^{-5}~\eVq}$
      & $7.42_{-0.20}^{+0.21}$ & $6.82 \to 8.04$
      & $7.42_{-0.20}^{+0.21}$ & $6.82 \to 8.04$
      \\[3mm]
      & $\dfrac{\Dmq_{3\ell}}{10^{-3}~\eVq}$
      & $+2.515_{-0.028}^{+0.028}$ & $+2.431 \to +2.599$
      & $+2.510_{-0.027}^{+0.027}$ & $+2.430 \to +2.593$\\[2mm]
      \hline\hline
    \end{tabular}
  \end{footnotesize}
  \caption{Normal Ordering NuFit~5.1 values~\cite{Esteban:2020cvm, nufit} for the neutrino observables.}
  \label{ta:NuFit51}
\end{table}


We note the significant differences between the two possibilities $n = 1+\sqrt{6}$ and $n = 1-\sqrt{6}$. This corresponds to a change of sign in the effective parameter $t$, which does not affect the predictions for $r$, $\theta_{13}$, $\theta_{12}$, but does affect the prediction for $\theta_{23}$ and $\delta$. This can be understood as the change of sign corresponds to changing from the tangent to a cotangent in the $\theta_{23}$ expression~\eqref{eq:theta23}, and for $\delta$~\eqref{eq:delta} to adding $\pi$.

While qualitatively both possibilities are similarly successful in reproducing the experimental data at $3 \sigma$, it is visible from the plots how the $1 \sigma$ range clearly favours different cases. It is worth emphasising how the new case we are considering is able to fit all observables at $1 \sigma$, with the exception of $\theta_{12}$, for which the $1 \sigma$ contour is just slightly above the intersection of all other observables, which include the very narrow contours from $\theta_{13}$ and from the mass ratio. To better quantify this we define
\begin{equation}
\chi^2 = \sum_i \left( \frac{ x_i^{\rm pred}-x_i^{\rm exp}}{ \sigma_i} \right)^2
\end{equation}
and list the respective $\chi^2$ values in Table~\ref{ta:chi-squareds}. For the $n=1+\sqrt 6$ case, $\chi^2 = 1.87$ can be obtained.
Table~\ref{ta:chi-squareds} also gives the predictions for $m_{ee}$ for the best-fit point in each case, where~\cite{Workman:2022ynf}:
\begin{equation}
m_{ee} = \left\lvert  \sum_i U_{ei}^2 m_i \right\rvert,
\end{equation}
which, in our case (since we are working in a basis where the charged-leptons are already diagonal, positive, and ordered) can be extracted simply from 
\begin{equation}
m_{ee} = \left\lvert \left(m_\nu\right)_{1,1} \right\rvert.
\end{equation}
From Eq.~\eqref{eq:mnu_mee}, we can see that this is identically $m_b$.
\begin{table}[h] 
\centering
\begin{footnotesize}
\begin{tabular}{|  c c c c c c c c c c c  |}
\hline \hline
\multicolumn{11}{|c|}{Goodness of fit against NuFit 5.1 values without SK atmospheric data}  \\
\hline
 $n$ & $\chi^2$ & r & $\beta/\pi$ & $m_b / 10^{-3}$ & $m_2^2 / 10^{-5}$ & $m_3^2/10^{-3}$ & $\theta_{12}$ & $\theta_{23}$ & $\theta_{13}$  & $\delta$ 	\\
\hline
  $1+\sqrt{6}$ 	& 29.47 	& 0.076	& 1.26 	& 2.33	& $7.19$	& $2.53$	& $34.29$	& $43.06$	& $8.78$	& $262$	\\
\hline
$1-\sqrt{6}$  	& 4.96	& 0.073	& 0.76	& 2.23	& $7.45$	& $2.51 $	& $34.34$	& $48.26$	& $8.55$	& $284$	\\
\hline
\hline
\multicolumn{11}{|c|}{Goodness of fit against NuFit 5.1 results with SK atmospheric data} \\
\hline
 $n$ & $\chi^2$ & r & $\beta/\pi$ & $m_b / 10^{-3}$ & $m_2^2 / 10^{-5}$ & $m_3^2/10^{-3}$ & $\theta_{12}$ & $\theta_{23}$ & $\theta_{13}$  & $\delta$ 	\\
\hline
  $1+\sqrt{6}$ 	& 1.87	& 0.074	& 1.24	& 2.30	& $7.42$	& $2.51 $	& $34.33$	& $42.03$	& $8.62$	& $257$ 	\\
\hline
 $1-\sqrt{6}$ 	& 25.79	& 0.077	& 0.74	& 2.33	& $7.15$	& $2.52 $	& $34.28$	& $46.76$	& $8.82$	& $277$ 	\\
\hline \hline
\end{tabular}
\caption{Our $\chi^2$ values for the different cases $n=1+\sqrt6$ and $n=1-\sqrt 6$. Note that from Eq.~\eqref{eq:mnu_mee} and the definition Eq.~\eqref{eq:mb_def}, the output parameter $m_{ee}$ is directly equal to the input parameters $m_b$.}
\label{ta:chi-squareds}
\end{footnotesize}
\end{table}

\section{Conclusion \label{sec:conc}}

In this paper, we have constructed the first complete model of the Littlest Modular Seesaw (LMS), 
based on CSD($1-\sqrt{6}$) $\approx$ CSD($-1.45$),
within a consistent framework based on 
multiple modular symmetries. We also proposed a new related possibility based on 
CSD($1+\sqrt{6}$) $\approx$ CSD($3.45$).
In each case, three $S_4$ modular symmetries are introduced, each with their respective modulus field at a distinct stabilizer,
leading to three separate residual subgroups, thus dispensing with vacuum alignment mechanisms. 
Of the three moduli, two are responsible implementing the viable  Littlest Seesaw leading to Trimaximal~1 mixing, which correlates non-trivially with the observed ratio of neutrino masses. The remaining modulus guarantees the charged lepton mass matrix is diagonal in the same basis, preserving the predictive power of the model.
The result, in the symmetry basis, is a diagonal charged lepton mass matrix and a LMS scenario of a particular kind.

Using a semi-analytical approach, we performed a $\chi^2$ analysis of each case case and showed that good agreement with neutrino oscillation data is obtained, for both possible octants of atmospheric angle,
including predictive relations between the leptonic mixing angles and the ratio of light neutrino masses, which non-trivially agree with the experimental values. It is noteworthy that in this very predictive setup, all the models fit the experimental data very well,
depending on the choice of stabilizers and data set, in one case to within approximately $1 \sigma$.
This is a remarkable achievement, given that the neutrino mass matrix in the diagonal charged lepton mass basis
is determined effectively by two real parameters, 
$m_a$, $m_b $ and one phase 
$\beta$ together with a discrete choice of $n=1\pm \sqrt{6}$.
For a given choice of $n$, the remaining three real parameters determine all the parameters in the neutrino sector,
namely all the neutrino masses and the entire PMNS matrix.

By extending the model to include a weighton and the double cover group $\Gamma'_4 \simeq S'_4$,
we are able to also account for the hierarchy of the charged leptons using modular symmetries,
without altering the neutrino predictions.

In summary, we have presented an extremely economical model of leptonic masses and mixing, by combining multiple modular symmetries with the littlest seesaw, and optionally adding a weighton. The latter accounts elegantly for the observed hierarchy of the lepton masses without the need for additional Froggatt-Nielsen style symmetries.

We argue that this is a minimal model of leptonic mixing, as we do not count the moduli as free continuous parameters given that we take them as stabilizers. As such, we have 3 real parameters in the charged lepton sector to fit the 3 masses, 1 real parameter that governs the overall neutrino mass scale, and just 2 effective parameters (the ratio $r=m_b/m_a$ and the phase $\beta$)
which fit the remaining observables: the neutrino mass ratio, the 3 PMNS mixing angles, the Dirac CP phase and a Majorana phase. The lightest neutrino mass is predicted to be zero and the PMNS phases are predicted in terms of the other observables.
Within this predictive setup we are able to fit all the neutrino oscillation data to within approximately $1 \sigma$.

\section*{Acknowledgements}
IdMV acknowledges funding from Funda\c{c}\~{a}o para a Ci\^{e}ncia e a Tecnologia (FCT) through the contract UID/FIS/00777/2020 and was supported in part by FCT through projects CFTP-FCT Unit 777 (UID/FIS/00777/2019), PTDC/FIS-PAR/29436/2017, CERN/FIS-PAR/0004/2019 and CERN/FIS-PAR/0008/2019 which are partially funded through POCTI (FEDER), COMPETE, QREN and EU.
 The work of ML is funded by
FCT Grant
No.PD/BD/150488/2019, in the framework of the Doctoral Programme
IDPASC-PT.  
SFK acknowledges the STFC Consolidated Grant ST/L000296/1 and the European Union's Horizon 2020 Research and Innovation programme under Marie Sklodowska-Curie grant agreement HIDDeN European ITN project (H2020-MSCA-ITN-2019//860881-HIDDeN).

\appendix

\section{Group Theory of $S_4$ \label{app:S4GT}}

In this appendix we summarize some relevant group theoretical details of $S_4$ (see \cite{deMedeirosVarzielas:2019cyj} and references therein).
The  products of irreps follows:
\begin{eqnarray}
&\mathbf{1^{\prime}}\otimes\mathbf{1^{\prime}}=\mathbf{1}, ~~\mathbf{1^{\prime}}\otimes\mathbf{2}=\mathbf{2}, ~~\mathbf{1^{\prime}}\otimes\mathbf{3}=\mathbf{3^{\prime}},~~ 
\mathbf{1^{\prime}}\otimes\mathbf{3^{\prime}}=\mathbf{3},\nonumber\\
&\mathbf{2}\otimes\mathbf{2}=\mathbf{1}\oplus\mathbf{1}^{\prime}\oplus\mathbf{2},~~
\mathbf{2}\otimes\mathbf{3}=\mathbf{2}\otimes\mathbf{3^{\prime}}=\mathbf{3}\oplus\mathbf{3}^{\prime},\nonumber\\
&\mathbf{3}\otimes\mathbf{3}=\mathbf{3^{\prime}}\otimes\mathbf{3^{\prime}}=\mathbf{1}\oplus \mathbf{2}\oplus\mathbf{3}\oplus\mathbf{3^{\prime}},~~
\mathbf{3}\otimes\mathbf{3^{\prime}}=\mathbf{1^{\prime}}\oplus \mathbf{2}\oplus\mathbf{3}\oplus\mathbf{3^{\prime}}\,.
\end{eqnarray}

In the basis we are using, the representation matrices for $T$, $S$ and $U$ are shown in Table \ref{tab:rep_matrix_main}.

\begin{table}[h!]
\begin{center}
\begin{footnotesize}
\begin{tabular}{|c|ccc|}
\hline\hline
   & $\rho(T)$ & $\rho(S)$ & $\rho(U)$  \\\hline
$\mathbf{1}$ & 1 & 1 & 1 \\
$\mathbf{1^{\prime}}$ & 1 & 1 & $-1$ \\
$\mathbf{2}$ & 
$\left(
\begin{array}{cc}
 \omega  & 0 \\
 0 & \omega ^2 \\
\end{array}
\right)$ & 
$\left(
\begin{array}{cc}
 1 & 0 \\
 0 & 1 \\
\end{array}
\right)$ & 
$\left(
\begin{array}{cc}
 0 & 1 \\
 1 & 0 \\
\end{array}
\right)$ \\

$\mathbf{3}$ &  $\left(
\begin{array}{ccc}
 1 & 0 & 0 \\
 0 & \omega ^2 & 0 \\
 0 & 0 & \omega  \\
\end{array}
\right)$ &
$\frac{1}{3} \left(
\begin{array}{ccc}
 -1 & 2 & 2 \\
 2 & -1 & 2 \\
 2 & 2 & -1 \\
\end{array}
\right)$ &
$\left(
\begin{array}{ccc}
 1 & 0 & 0 \\
 0 & 0 & 1 \\
 0 & 1 & 0 \\
\end{array}
\right)$ \\

$\mathbf{3^{\prime}}$ &  $\left(
\begin{array}{ccc}
 1 & 0 & 0 \\
 0 & \omega ^2 & 0 \\
 0 & 0 & \omega  \\
\end{array}
\right)$ &
$\frac{1}{3} \left(
\begin{array}{ccc}
 -1 & 2 & 2 \\
 2 & -1 & 2 \\
 2 & 2 & -1 \\
\end{array}
\right)$ &
$-\left(
\begin{array}{ccc}
 1 & 0 & 0 \\
 0 & 0 & 1 \\
 0 & 1 & 0 \\
\end{array}
\right)$ \\ \hline\hline

\end{tabular}
\caption{\label{tab:rep_matrix_main} In the basis used, the representation matrices for $T$, $S$ and $U$, with $\omega=e^{2\pi i/3}$.}
\end{footnotesize}
\end{center}
\end{table}
%

In this basis, the product of 3 dimensional irreps $a$ and $b$:
\begin{eqnarray}
(ab)_\mathbf{1_i} &=& a_1b_1 + a_2b_3 + a_3b_2 \,,\nonumber\\
(ab)_\mathbf{2} &=& (a_2b_2 + a_1b_3 + a_3b_1,~ a_3b_3 + a_1b_2 + a_2b_1)^T \,,\nonumber\\
(ab)_{\mathbf{3_i}} &=& (2a_1b_1-a_2b_3-a_3b_2, 2a_3b_3-a_1b_2-a_2b_1, 2a_2b_2-a_3b_1-a_1b_3)^T \,,\nonumber\\
(ab)_{\mathbf{3_j}} &=& (a_2b_3-a_3b_2, a_1b_2-a_2b_1, a_3b_1-a_1b_3)^T \,,
\label{eq:CG2}
\end{eqnarray}
for
\begin{eqnarray} \label{eq:_i_j}
&&\mathbf{1_i}=\mathbf{1}\,, ~\; \mathbf{3_i}=\mathbf{3}\,, ~\; \mathbf{3_j}=\mathbf{3'}\,~\; \text{for} ~\; a\sim b \sim \mathbf{3}\,,~ \mathbf{3^\prime} \,, \nonumber\\
&&\mathbf{1_i}=\mathbf{1'}\,,~  \mathbf{3_i}=\mathbf{3'}\,,~ \mathbf{3_j}=\mathbf{3}\,~\;\; \text{for} ~\; a\sim  \mathbf{3}\,,~ b \sim \mathbf{3'}\,.
\end{eqnarray} 

The expressions for the product of 2 dimensional irreps $a=(a_1, a_2)^T$ and $b=(b_1, b_2)^T$ are:
\begin{eqnarray}
(ab)_\mathbf{1} &=& a_1b_2 + a_2b_1 \,,\quad
(ab)_\mathbf{1^\prime} = a_1b_2 - a_2b_1 \,,\quad
(ab)_{\mathbf{2}} = (a_2b_2, a_1b_1)^T \,.
\label{eq:CG_doublets}
\end{eqnarray}

\section{Stabilizers and Residual symmetry  \label{app:stabs}}

In the basis we work in, we can make the following mapping of modular generators~\cite{deMedeirosVarzielas:2019cyj}:
\begin{equation}\label{eq:basis}
S=T_\tau^2, \qquad T=S_\tau T_\tau, \qquad U = T_\tau S_\tau T_\tau^2 S_\tau, 
\end{equation}
where $S_\tau$ and $T_\tau$ are the usual modular generators of $\Gamma_N$:
\begin{equation}
S_\tau = \begin{pmatrix} 0 & 1 \\ -1 & 0 \end{pmatrix}, \qquad T_\tau = \begin{pmatrix} 1 & 1 \\ 0 & 1 \end{pmatrix}
\end{equation}
which act on the modulus field as
\begin{equation}
\gamma \tau = \dfrac{a \tau + b}{c \tau +d}, \qquad \gamma = \begin{pmatrix} a & b \\ c & d \end{pmatrix}.
\end{equation}
With the requirement that $\tau=\tau+4$, which must hold true for $\Gamma_4$, we can compute the corresponding $\gamma$ for $U$ and $SU$~\cite{deMedeirosVarzielas:2019cyj}:
\begin{equation}
\gamma(U) =\begin{pmatrix} 1 & -1 \\ 2 & -1 \end{pmatrix} , \qquad \gamma(SU) = \begin{pmatrix} 5 & -3 \\ 2 & -1 \end{pmatrix}.
\end{equation}
Now, due to $T_\tau^4=\mathbf{1}$, the choice of $\gamma(g)$ is not unique. Indeed, any element of $S_4$, $\gamma(g)$: 
\begin{equation}
\gamma(g)= \begin{pmatrix} a & b \\ c & d \end{pmatrix}, \qquad ad-bc=1, \quad a,b,c,d \in \mathbb{Z},
\end{equation}
is equivalent to 
\begin{equation}
\gamma'(g) = (\pm 1) \begin{pmatrix} 4 k_a +a & 4 k_b +b \\ 4 k_c +c & 4 k_d +d \end{pmatrix}, \qquad 4 k_a k_d + a k_d + d k_a = 4 k_b k_c b k_c + c k_d, \quad k_x \in \mathbb{Z}
\end{equation}
where the constraint comes from requiring that $\gamma'(g)$ also satisfies $ad-bc=1$.

By choosing the following sets of integers, we arrive at equivalent representations of the $\gamma(U)$ and $\gamma(SU)$ matrices:
\begin{eqnarray}
\gamma_1(U) &= \begin{pmatrix} 1 & -1 \\ 2 & -1 \end{pmatrix} &\equiv \gamma(U), \\
\gamma_2(U) &= \begin{pmatrix} -3 & -5 \\ 2 & 3 \end{pmatrix}, &\quad \begin{matrix} k_a = -1 & k_b = -1 & k_c = 0 & k_d = 1 \end{matrix}, \\
\gamma_1(SU) &= \begin{pmatrix} -1 & -1 \\ 2 & 1 \end{pmatrix}, &\quad \begin{matrix} k_a = -1 & k_b = 1 & k_c = -1 & k_d = 0 \end{matrix}, \\
\gamma_2(SU) &=\begin{pmatrix} -3 & 5 \\ -2 & 3 \end{pmatrix}, &\quad \begin{matrix} k_a = -2 & k_b = 2 & k_c = -1 & k_d = 1 \end{matrix}.
\end{eqnarray}
Using these matrices, it is straightforward to show that
\begin{eqnarray}
\gamma_1(U) \tau_A &=& \tau_A , \quad \tau_A = \dfrac{1+i}{2}\\
\gamma_2(U) \tau'_A &=& \tau'_A, \quad \tau'_A = \dfrac{-3+i}{2} \\
\gamma_1(SU) \tau_B &=& \tau_B, \quad \tau_B = \dfrac{-1+i}{2} \\
\gamma_2(SU) \tau_B &=& \tau_B, \quad \tau_B = \dfrac{3+i}{2} .
\end{eqnarray}
In other words, $\tau_A$ and $\tau'_A$ are stabilisers of the modular generator $U$, and that $\tau_B$ (either version) is a stabiliser of the modular generator $SU$ in our chosen basis.

To further corroborate that the stabilisers are leaving an unbroken subgroup, we can check that the respective modular forms are eigenvectors of the appropriate representation matrices.
From Appendix~\ref{app:S4GT}, we have
\begin{equation}
\rho_\mathbf{3'}(S) = \dfrac{1}{3}\begin{pmatrix} -1 & 2 & 2 \\ 2 & -1 & 2 \\ 2 & 2 &-1 \end{pmatrix}, \qquad \rho_\mathbf{3'}(U) = -\begin{pmatrix} 1 & 0 & 0 \\ 0 & 0 & 1 \\ 0 & 1 &0 \end{pmatrix}, \qquad \rho_\mathbf{3'}(SU) = \dfrac{1}{3}\begin{pmatrix} 1 & -2 & -2 \\ -2 & -2 & 1 \\ -2 & 1 &-2 \end{pmatrix},
\end{equation}
from which is straightforward to arrive at
\begin{equation}
\rho_\mathbf{3'}(U) \cdot \begin{pmatrix} 0 \\ -1 \\ 1 \end{pmatrix} = (+1) \begin{pmatrix} 0 \\ -1 \\ 1 \end{pmatrix}, \qquad \rho_\mathbf{3'}(SU) \cdot \begin{pmatrix} 1 \\ 1\pm\sqrt{6} \\ 1\mp\sqrt{6} \end{pmatrix} = (-1) \begin{pmatrix} 1 \\ 1\pm\sqrt{6} \\ 1\mp\sqrt{6} \end{pmatrix}, 
\end{equation}
agreeing with the expected results. We note that both modular forms $\begin{pmatrix} 1 & 1\pm \sqrt{6} & 1 \mp \sqrt{6} \end{pmatrix}$ have an eigenvalue $-1$, which is a consequence of~\cite{deMedeirosVarzielas:2019cyj}
\begin{equation}
(c \tau + d)^{-2k} = (2 \tau_{SU} +1)^{-2k} = (-1)^k,
\end{equation}
where $k=1$ for $Y_{\mathbf{3'}}^{(2)}$. As such, we only preserve a residual flavour symmetry $U$ by the modular form of $\tau_A$, whereas the modular forms of $\tau_B$ only preserve a residual modular symmetry.

\section{Weighton models \label{app:weighton}}

\subsection{A minimal weighton model \label{sec:weighton}}

We now modify the model presented in the main text to include a weighton field $\phi$. In order to preserve the features of the previous model (particularly the diagonal charged-lepton mass matrix) we employ ${S'}_4$ modular symmetries~\cite{Novichkov:2020eep} instead of $S_4$. 

The assignments of the fields under the symmetries are listed in Table~\ref{ta:model3}.
Notice that this implementation of the weighton is distinguished from the standard one as the weighton is assigned to non-trivial representations of ${S'}_4^A$, ${S'}_4^B$, and ${S'}_4^C$. Due to this and the representations of the charged leptons, the invariant terms have the desired modular forms $Y_\tau$, $Y_\mu$ and $Y_e$ respectively for the field combinations $L \tau^c$, $L \mu^c \phi$ and $L e^c \phi^3$. This is shown (in green) in Table~\ref{ta:WeightonRepsv2}, where other possibilities are not invariant.

Since there are no charged leptons with weights under ${S'}_4^{A,B}$, the charged-leptons Yukawa modular forms must be singlets under ${S'}_4^{A,B}$ with weight 0 under these symmetries.

By having chosen the weighton to have a positive weight under ${S'}_4^{C}$, there are no additional contributions beyond the leading order ones, as the Yukawa modular forms also have positive weight.
An alternative solution, where the weighton has a negative weight under ${S'}_4^{C}$, is presented in Appendix~\ref{app:negative}.

\begin{table}[h]
\centering
\begin{footnotesize}
 \begin{tabular}{| l | c c c c c c|}
\hline \hline
Field & ${S'}_4^A$ & ${S'}_4^B$ & ${S'}_4^C$ & \!$2k_A$\! & \!$2k_B$\! & \!$2k_C$\!\\ 
\hline \hline
$L$ & $\mathbf{1}$ & $\mathbf{1}$ & $\mathbf{3}$ & 0 & 0 & 0\\
$e^c$ & $\mathbf{\hat{1}}$ & $\mathbf{\hat{1}}$ & $\mathbf{1}'$ & 0 & 0 & \!$-12$\! \\
$\mu^c$ & $\mathbf{\hat{1}'}$ & $\mathbf{\hat{1}'}$ & $\mathbf{1}'$ & 0 & 0 & \!$-6$\! \\
$\tau^c$ & $\mathbf{1}$ & $\mathbf{1}$ & $\mathbf{1}'$ & 0 & 0 & \!$-2$\! \\
$N_A^c$ & $\mathbf{1'}$ & $\mathbf{1}$ & $\mathbf{1}$ & $-4$ & 0 & 0 \\
$N_B^c$ & $\mathbf{1}$ & $\mathbf{1'}$ & $\mathbf{1}$ & 0 & $-2$ & 0 \\
\hline 
$\Phi_{AC}$ & $\mathbf{3}$ & $\mathbf{1}$ & $\mathbf{3}$ & 0 & 0 & 0 \\
$\Phi_{BC}$ & $\mathbf{1}$ & $\mathbf{3}$ & $\mathbf{3}$ & 0 & 0 & 0 \\
\hline
$\phi$ & $\mathbf{\hat{1}}$ & $\mathbf{\hat{1}} $ & $ \mathbf{\hat{1}}$ & 0 & 0 & \!$+2$\! \\
\hline \hline
\end{tabular}
\begin{tabular}{| l | c c c c c c|}
\hline \hline
Yuk/Mass &${S'}_4^A$ & ${S'}_4^B$ & ${S'}_4^C$ & \!$2k_A$\! & \!$2k_B$\! & \!$2k_C$\!\\
\hline \hline
$Y_e(\tau_C) $ & $\mathbf{1}$ & $\mathbf{1}$ & $\mathbf{3}'$ & 0 & 0 & $6$ \\
$Y_\mu(\tau_C)$ & $\mathbf{1}$ & $\mathbf{1}$ & $\mathbf{3}'$ & 0 & 0 & $4$ \\
$Y_\tau(\tau_C)$ & $\mathbf{1}$ & $\mathbf{1}$ & $\mathbf{3}'$ & 0 & 0 & $2$ \\
$Y_A(\tau_A)$ & $\mathbf{3'}$ & $\mathbf{1}$ & $\mathbf{1}$ & $4$ & 0 & 0 \\
$Y_B(\tau_B)$ & $\mathbf{1}$ & $\mathbf{3'}$ & $\mathbf{1}$ & 0 & $2$ & 0 \\\hline
$M_A(\tau_A)$ & $\mathbf{1}$ & $\mathbf{1}$ & $\mathbf{1}$ & $8$ & 0 & 0 \\
$M_B(\tau_B)$ & $\mathbf{1}$ & $\mathbf{1}$ & $\mathbf{1}$ & 0 & $4$ & 0 
\\
\hline \hline
\end{tabular}
\caption{Assignments of fields for the weighton version of the model.}
\label{ta:model3}
\end{footnotesize}
\end{table}

\begin{table}[h!]
\centering
\begin{footnotesize}
\begin{tabular}{| l | c c c c c |}
\hline \hline 
	& $\phi^{0}$ &  $\phi^{1}$ &  $\phi^{2}$ &  $\phi^{3}$ &  $\phi^{4}$  \\
\hline \hline 
$L e^c$    
		& $\left(\textcolor{black}{\mathbf{\hat{1}}_{0},  \mathbf{\hat{1}}_{0}}, \mathbf{\hat{3}'}_{-12} \right)$ & 
		$\left( \textcolor{black}{\mathbf{1'}_{0},  \mathbf{1'}_{0}},\mathbf{3}_{-10}\right)$ & 
		$\left( \textcolor{black}{\mathbf{\hat{1}'}_{0}, \mathbf{\hat{1}'}_{0}}, \mathbf{\hat{3}}_{-8}\right)$ & 
		\color{ForestGreen}{$\left( \mathbf{1}_{0}, \mathbf{1}_{0},  \mathbf{3'}_{-6}\right)$} 	&
		 $\left(\textcolor{black}{\mathbf{\hat{1}}_{0},  \mathbf{\hat{1}}_{0}}, \mathbf{\hat{3}'}_{-4} \right)$\\
$L \mu^c$    
		& $\left(  \textcolor{black}{\mathbf{\hat{1}'}_{0}, \mathbf{\hat{1}'}_{0}},  \mathbf{\hat{3}}_{-6}\right)$ & 
		\color{ForestGreen}{$\left( \mathbf{1}_{0}, \mathbf{1}_{0}, \mathbf{3}'_{-4}\right)$} & 
		$\left(  \textcolor{black}{\mathbf{\hat{1}}_{0}, \mathbf{\hat{1}}_{0}},  \mathbf{\hat{3}'}_{-2}\right)$ & 
		$\left(  \textcolor{black}{\mathbf{1'}_{0}, \mathbf{1'}_{0}},  \mathbf{3}_{0}\right)$ & 
		$\left( \mathbf{\hat{1}'}_{0},  \mathbf{\hat{1}'}_{0}, \textcolor{black}{\mathbf{\hat{3}}_{+2}}\right)$  \\
$L \tau^c$    
		& \color{ForestGreen}{$\left( \mathbf{1}_{0}, \mathbf{1}_{0}, \mathbf{3'}_{-2}\right)$} & 
		$\left(  \mathbf{\textcolor{black}{\hat{1}_{0}}, \mathbf{\hat{1}}_{0}}, \mathbf{\hat{3}'}_{0}\right)$ & 
		$\left(  \mathbf{1'}_{0}, \mathbf{1'}_{0}, \textcolor{black}{\mathbf{3}_{+2}}\right)$ & 
		$\left(  \mathbf{\hat{1}'}_{0}, \mathbf{\hat{1}'}_{0}, \textcolor{black}{\mathbf{\hat{3}}_{+4}}\right)$  &
		$\left( \mathbf{1}_{0}, \mathbf{1}_{0}, \textcolor{black}{\mathbf{3'}_{+6}}\right)$ \\
\hline \hline
\end{tabular}
\caption{Irreps of the leptonic tensor products with different powers of the weighton. The invariant combinations are highlighted in green.}
\label{ta:WeightonRepsv2}
\end{footnotesize}
\end{table}

\subsection{An alternative weighton model \label{app:negative}}

In this subsection we provide an alternative weighton model, that does not require assigning large modular weights to the charged lepton fields.

This allows fields (in particular charged lepton fields) to be assigned as distinct non-trivial singlets of the underlying modular symmetries, as shown in Table~\ref{ta:model2}.

\begin{table}[h!] 
\centering
\begin{footnotesize}
\begin{tabular}{| l | c c c c c c|}
\hline \hline
Field & ${S'}_4^A$ & ${S'}_4^B$ & ${S'}_4^C$ & \!$2k_A$\! & \!$2k_B$\! & \!$2k_C$\!\\ 
\hline \hline
$L$ & $\mathbf{1}$ & $\mathbf{1}$ & $\mathbf{3}$ & 0 & 0 & 0\\
$e^c$ & $\mathbf{\hat{1}}$ & $\mathbf{\hat{1}}$ & $\mathbf{1}'$ & 0 & 0 & \!$0$\! \\
$\mu^c$ & $\mathbf{\hat{1}'}$ & $\mathbf{\hat{1}'}$ & $\mathbf{1}'$ & 0 & 0 & \!$-2$\! \\
$\tau^c$ & $\mathbf{1}$ & $\mathbf{1}$ & $\mathbf{1}'$ & 0 & 0 & \!$-2$\! \\
$N_A^c$ & $\mathbf{1'}$ & $\mathbf{1}$ & $\mathbf{1}$ & $-4$ & 0 & 0 \\
$N_B^c$ & $\mathbf{1}$ & $\mathbf{1'}$ & $\mathbf{1}$ & 0 & $-2$ & 0 \\
\hline 
$\Phi_{AC}$ & $\mathbf{3}$ & $\mathbf{1}$ & $\mathbf{3}$ & 0 & 0 & 0 \\
$\Phi_{BC}$ & $\mathbf{1}$ & $\mathbf{3}$ & $\mathbf{3}$ & 0 & 0 & 0 \\
\hline
$\phi$ & $\mathbf{\hat{1}}$ & $\mathbf{\hat{1}} $ & $ \mathbf{\hat{1}}$ & 0 & 0 & \!$-2$\! \\
\hline \hline
\end{tabular}
\begin{tabular}{| l | c c c c c c|}
\hline \hline
Yuk/Mass &${S'}_4^A$ & ${S'}_4^B$ & ${S'}_4^C$ & \!$2k_A$\! & \!$2k_B$\! & \!$2k_C$\!\\
\hline \hline
$Y_e(\tau_C)$ & $\mathbf{1}$ & $\mathbf{1}$ & $\mathbf{3}'$ & 0 & 0 & $6$ \\
$Y_\mu(\tau_C)$ & $\mathbf{1}$ & $\mathbf{1}$ & $\mathbf{3}'$ & 0 & 0 & $4$ \\
$Y_\tau(\tau_C)$ & $\mathbf{1}$ & $\mathbf{1}$ & $\mathbf{3}'$ & 0 & 0 & $2$ \\
$Y_A(\tau_A)$ & $\mathbf{3'}$ & $\mathbf{1}$ & $\mathbf{1}$ & $4$ & 0 & 0 \\
$Y_B(\tau_B)$ & $\mathbf{1}$ & $\mathbf{3'}$ & $\mathbf{1}$ & 0 & $2$ & 0 \\\hline
$M_A(\tau_A)$ & $\mathbf{1}$ & $\mathbf{1}$ & $\mathbf{1}$ & $8$ & 0 & 0 \\
$M_B(\tau_B)$ & $\mathbf{1}$ & $\mathbf{1}$ & $\mathbf{1}$ & 0 & $4$ & 0 
\\
\hline \hline
\end{tabular}
\caption{Assignments of fields for the alternative weighton version of the model.}
\label{ta:model2}
\end{footnotesize}
\end{table}

\begin{table}[h!]
\centering
\begin{footnotesize}
\begin{tabular}{| l | c c c c c |}
\hline \hline 
	& $\phi^{0}$ &  $\phi^{1}$ &  $\phi^{2}$ &  $\phi^{3}$ &  $\phi^{4}$  \\
\hline \hline 
$L e^c$    
		& $\left(\textcolor{black}{\mathbf{\hat{1}}_{0}},  \mathbf{\hat{1}}_{0}, \mathbf{\hat{3}'}_{0} \right)$ & 
		$\left( \textcolor{black}{\mathbf{1'}_{0}},  \mathbf{1'}_{0},\mathbf{3}_{-2}\right)$ & 
		$\left( \textcolor{black}{\mathbf{\hat{1}'}_{0}}, \mathbf{\hat{1}'}_{0}, \mathbf{\hat{3}}_{-4}\right)$ & 
		\color{ForestGreen}{$\left( \mathbf{1}_{0}, \mathbf{1}_{0},  \mathbf{3'}_{-6}\right)$} 	&
		 $\left(\textcolor{black}{\mathbf{\hat{1}}_{0}},  \mathbf{\hat{1}}_{0}, \mathbf{\hat{3}'}_{-8} \right)$\\
$L \mu^c$    
		& $\left(  \textcolor{black}{\mathbf{\hat{1}'}_{0}}, \mathbf{\hat{1}'}_{0},  \mathbf{\hat{3}}_{-2}\right)$ & 
		\color{ForestGreen}{$\left( \mathbf{1}_{0}, \mathbf{1}_{0}, \mathbf{3}'_{-4}\right)$} & 
		$\left(  \textcolor{black}{\mathbf{\hat{1}}_{0}}, \mathbf{\hat{1}}_{0},,  \mathbf{\hat{3}'}_{-6}\right)$ & 
		$\left(  \textcolor{black}{\mathbf{1'}_{0}}, \mathbf{1'}_{0},  \mathbf{3}_{-8}\right)$ & 
		$\left( \textcolor{black}{\mathbf{\hat{1}'}_{0}},  \mathbf{\hat{1}'}_{0}, \mathbf{\hat{3}}_{-10}\right)$  \\
$L \tau^c$    
		& \color{ForestGreen}{$\left( \mathbf{1}_{0}, \mathbf{1}_{0}, \mathbf{3'}_{-2}\right)$} & 
		$\left(  \mathbf{\textcolor{black}{\hat{1}_{0}}}, \mathbf{\hat{1}}_{0}, \mathbf{\hat{3}'}_{-4}\right)$ & 
		$\left(  \textcolor{black}{\mathbf{1'}_{0}}, \mathbf{1'}_{0}, \mathbf{3}_{-6}\right)$ & 
		$\left(  \textcolor{black}{\mathbf{\hat{1}'}_{0}}, \mathbf{\hat{1}'}_{0}, \mathbf{\hat{3}}_{-8}\right)$  &
		\color{ForestGreen}{$\left( \mathbf{1}_{0}, \mathbf{1}_{0}, \mathbf{3'}_{-10}\right)$} \\
\hline 
$L \Phi_{AC} N_A^c$    
		& \color{ForestGreen}{$\left(\mathbf{3'}_{-4},  \mathbf{1}_{0}, \mathbf{1}_{0}\right)$} & 
						$\left( \mathbf{\hat{3}'}_{-4}, \textcolor{black}{\mathbf{\hat{1}}_{0}},  \mathbf{\hat{1}}_{-2}\right)$ & 
						$\left(  \mathbf{3}_{-4}, \textcolor{black}{\mathbf{1'}_{0}}, \mathbf{1'}_{-4}\right)$ & 
						$\left( \mathbf{\hat{3}}_{-4},  \textcolor{black}{\mathbf{\hat{1}'}_{0}}, \mathbf{\hat{1}'}_{-6}\right)$  &
		\color{ForestGreen}{$\left(\mathbf{3'}_{-4},  \mathbf{1}_{0}, \mathbf{1}_{-8}\right)$} \\
$L \Phi_{BC} N_B^c$    
		& \color{ForestGreen}{$\left(\mathbf{1}_{0},  \mathbf{3'}_{-2}, \mathbf{1}_{0}\right)$} & 
						$\left(\textcolor{black}{\mathbf{\hat{1}}_{0}},  \mathbf{\hat{3}'}_{-2},  \mathbf{\hat{1}}_{-2}\right)$ & 
						$\left(\textcolor{black}{ \mathbf{1'}_{0}},  \mathbf{3}_{-2},  \mathbf{1'}_{-4}\right)$ & 
						$\left(\textcolor{black}{ \mathbf{\hat{1}'}_{0}}, \mathbf{\hat{3}}_{-2},  \mathbf{\hat{1}'}_{-6}\right)$  &
		\color{ForestGreen}{$\left( \mathbf{1}_{0},  \mathbf{3'}_{-2}, \mathbf{1}_{-8}\right)$} \\
\hline 
$N_A^c N_A^c$    
		& \color{ForestGreen}{$\left( \mathbf{1}_{-8},\mathbf{1}_{0},  \mathbf{1}_{0}\right)$} & 
						$\left(\mathbf{\hat{1}}_{-8},  \textcolor{black}{\mathbf{\hat{1}}_{0}}, \mathbf{\hat{1}}_{-2} \right)$ & 
						$\left( \mathbf{1'}_{-8},  \textcolor{black}{\mathbf{1'}_{0}},\mathbf{1'}_{-4} \right)$ & 
						$\left(\mathbf{\hat{1}'}_{-8},  \textcolor{black}{\mathbf{\hat{1}'}_{0}}, \mathbf{\hat{1}'}_{-6} \right)$ & 
		 \color{ForestGreen}{$\left(  \mathbf{1}_{-8}, \mathbf{1}_{0},\mathbf{1}_{-8}\right)$} \\
$N_B^c N_B^c$    
		& \color{ForestGreen}{$\left( \mathbf{1}_{0}, \mathbf{1}_{-4}, \mathbf{1}_{0}\right)$} & 
						$\left( \textcolor{black}{\mathbf{\hat{1}}_{0}},  \mathbf{\hat{1}}_{-4},\mathbf{\hat{1}}_{-2} \right)$ & 
						$\left(  \textcolor{black}{\mathbf{1'}_{0}},\mathbf{1'}_{-4}, \mathbf{1'}_{-4} \right)$ & 
						$\left( \textcolor{black}{\mathbf{\hat{1}'}_{0}},  \mathbf{\hat{1}'}_{-4},\mathbf{\hat{1}'}_{-6} \right)$ & 
		 \color{ForestGreen}{$\left( \mathbf{1}_{0},  \mathbf{1}_{-4},\mathbf{1}_{-8}\right)$} \\
$N_A^c N_B^c$    
		& 				$\left( \mathbf{1'}_{-4}, \textcolor{black}{\mathbf{1'}_{-2}}, \mathbf{1}_{0}\right)$ & 
						$\left( \mathbf{\hat{1}'}_{-4}, \textcolor{black}{\mathbf{\hat{1'}}_{-2}}, \mathbf{\hat{1}}_{-2} \right)$ & 
						$\left( \mathbf{1}_{-4}, \textcolor{black}{\mathbf{1}_{-2}}, \mathbf{1'}_{-4} \right)$ & 
						$\left( \mathbf{\hat{1}}_{-4}, \textcolor{black}{\mathbf{\hat{1}}_{-2}}, \mathbf{\hat{1}'}_{-6} \right)$ & 
		  				$\left( \mathbf{1'}_{-4}, \textcolor{black}{\mathbf{1'}_{-2}}, \mathbf{1}_{-8}\right)$   \\
\hline 
$N_A^c \Phi_{AC} N_A^c$    
			& 			$\left( \mathbf{3}_{-8},\mathbf{1}_{0},  \textcolor{black}{\mathbf{3}_{0}}\right)$ & 
						$\left(\mathbf{\hat{3}}_{-8},  \textcolor{black}{\mathbf{\hat{1}}_{0}}, \mathbf{\hat{3}}_{-2} \right)$ & 
						$\left( \mathbf{3'}_{-8},  \textcolor{black}{\mathbf{1'}_{0}},\mathbf{3'}_{-4} \right)$ & 
						$\left(\mathbf{\hat{3}'}_{-8},  \textcolor{black}{\mathbf{\hat{1}'}_{0}}, \mathbf{\hat{3}'}_{-6} \right)$ & 
		 \color{ForestGreen}{$\left(  \mathbf{3}_{-8}, \mathbf{1}_{0},\mathbf{3}_{-8}\right)$ }\\
$N_B^c \Phi_{AC} N_B^c$    
		& 				$\left( \textcolor{black}{\mathbf{3}_{0}}, \mathbf{1}_{-4}, \mathbf{3}_{0}\right)$ & 
						$\left( \textcolor{black}{\mathbf{\hat{3}}_{0}},  \mathbf{\hat{1}}_{-4},\mathbf{\hat{3}}_{-2} \right)$ & 
						$\left(  \textcolor{black}{\mathbf{3'}_{0},\mathbf{1'}_{-4}}, \mathbf{3'}_{-4} \right)$ & 
						$\left( \textcolor{black}{\mathbf{\hat{3}'}_{0}},  \mathbf{\hat{1}'}_{-4},\mathbf{\hat{3}'}_{-6} \right)$ & 
						$\left( \textcolor{black}{\mathbf{3}_{0}},  \mathbf{1}_{-4},\mathbf{3}_{-8}\right)$ \\
$N_A^c \Phi_{AC} N_B^c$    
		& 				$\left( \mathbf{3'}_{-4}, \textcolor{black}{\mathbf{1'}_{-2}}, \mathbf{3}_{0}\right)$ & 
						$\left( \mathbf{\hat{3}'}_{-4}, \textcolor{black}{\mathbf{\hat{1}'}_{-2}}, \mathbf{\hat{3}}_{-2} \right)$ & 
						$\left( \mathbf{3}_{-4}, \textcolor{black}{\mathbf{1}_{-2}}, \mathbf{3'}_{-4} \right)$ & 
						$\left( \mathbf{\hat{3}}_{-4}, \textcolor{black}{\mathbf{\hat{1}}_{-2}}, \mathbf{\hat{3}'}_{-6} \right)$ & 
		  				$\left( \mathbf{3'}_{-4}, \textcolor{black}{\mathbf{1'}_{-2}}, \mathbf{3}_{-8}\right)$   \\
\hline 
$N_A^c \Phi_{BC} N_A^c$    
		& 				$\left( \mathbf{1}_{-8},\textcolor{black}{\mathbf{3}_{0}},  \mathbf{3}_{0}\right)$ & 
						$\left(\mathbf{\hat{1}}_{-8},  \textcolor{black}{\mathbf{\hat{3}}_{0}}, \mathbf{\hat{3}}_{-2} \right)$ & 
						$\left( \mathbf{1'}_{-8},  \textcolor{black}{\mathbf{3'}_{0}},\mathbf{3'}_{-4} \right)$ & 
						$\left(\mathbf{\hat{1}'}_{-8},  \textcolor{black}{\mathbf{\hat{3}'}_{0}}, \mathbf{\hat{3}'}_{-6} \right)$ & 
		 				$\left(  \mathbf{1}_{-8}, \textcolor{black}{\mathbf{3}_{0}},\mathbf{3}_{-8}\right)$ \\
$N_B^c \Phi_{BC} N_B^c$    
		& 				$\left( \mathbf{1}_{0}, \mathbf{3}_{-4}, \textcolor{black}{\mathbf{3}_{0}}\right)$ & 
						$\left( \textcolor{black}{\mathbf{\hat{1}}_{0}},  \mathbf{\hat{3}}_{-4},\mathbf{\hat{3}}_{-2} \right)$ & 
						$\left(  \textcolor{black}{\mathbf{1'}_{0}},\mathbf{3'}_{-4}, \mathbf{3'}_{-4} \right)$ & 
						$\left( \textcolor{black}{\mathbf{\hat{1}'}_{0}},  \mathbf{\hat{3}'}_{-4},\mathbf{\hat{3}'}_{-6} \right)$ & 
		 				\color{ForestGreen}{$\left( \mathbf{1}_{0},  \mathbf{3}_{-4},\mathbf{3}_{-8}\right)$ }\\
$N_A^c \Phi_{BC} N_B^c$    
		& 				$\left( \mathbf{1'}_{-4}, \mathbf{3'}_{-2}, \textcolor{black}{\mathbf{3}_{0}}\right)$ & 
						$\left( \textcolor{black}{\mathbf{\hat{1}'}_{-4}}, \mathbf{\hat{3}'}_{-2}, \mathbf{\hat{3}}_{-2} \right)$ & 
						$\left( \mathbf{1}_{-4}, \textcolor{black}{\mathbf{3}_{-2}}, \mathbf{3'}_{-4} \right)$ & 
						$\left( \mathbf{\hat{1}}_{-4}, \textcolor{black}{\mathbf{\hat{3}}_{-2}}, \mathbf{\hat{3}'}_{-6} \right)$ & 
		  				\color{ForestGreen}{$\left( \mathbf{1'}_{-4}, \mathbf{3'}_{-2}, \mathbf{3}_{-8}\right)$ }  \\
\hline \hline
\end{tabular}
\caption{Irreps of the leptonic tensor products with different powers of the weighton following the new assignments. The invariant combinations are highlighted in green.}
\label{ta:WeightonReps}
\end{footnotesize}
\end{table}

Table~\ref{ta:WeightonReps} shows the assignments of the different field combinations and clarifies how the non-trivial singlet choices of the charged leptons allow only one coupling at leading order of powers of $\phi$, with the next leading order term appearing only with the insertion of additional $\phi^4$ \footnote{Since the weighton is charged under ${S'}^C_4$, and the 1D irreps have at most $\mathbf{r}^4=\mathbf{1}$, there will always be corrections to the leading terms with 4 more weighton insertions. This is avoided by taking the weighton model of Appendix~\ref{sec:weighton}.}. We estimate this suppression factor should to be around $10^{-5}$ by assuming $\mathcal{O}(1)$ couplings for the charged leptons \footnote{Namely, we take $\langle \phi \rangle /M = \epsilon = 6.5 \times 10^{-2}$, to have $m_\mu  \sim 0.92\, \epsilon\, m_\tau$ and $m_e  \sim 1.08\, \epsilon^3\, m_\tau$.}.

\bibliographystyle{JHEP}
\bibliography{LMS.bib}

\end{document}